# Model Checking of BPMN Models for Reconfigurable Workflows

Juan Carlos Polanco Aguilar[1], Koji Hasebe[1], Manuel Mazzara[2], and Kazuhiko Kato[1]

[1]Graduate School of Systems and Information Engineering, University of Tsukuba
[2]School of Computing Science, Newcastle University, UK

*Abstract*—Nowadays, business enterprises often need to dynamically reconfigure their internal processes in order to improve the efficiency of the business flow. However, modifications of the workflow usually lead to several problems in terms of deadlock freedom, completeness and security. A solid solution to these problems consists in the application of model checking techniques in order to verify if specific properties of the workflow are preserved by the change in configuration. Our goal in this work is to develop a formal verification procedure to deal with these problems. The first step consists in developing a formal definition of a BPMN model of a business workflow. Then, a given BPMN model is translated into a formal model specified in Promela. Finally, by using the SPIN model checker, the correctness of the reconfigured workflow is verified.

## I. Purpose

Web services have recently seen a rapid growth of their usability and functionality. The advantages that this technology can provide to the business framework are also growing fast. The interaction from different services allow the user to discover and utilize a better transactional side of the Internet. Web services also provides several benefits such as interoperability and reusability across platforms, applications and programming languages by the introduction of standards and integration profiles.

Although this technology has provided many advantages, some of the services still require to be flexible and available for different changes in their business logic. Reconfiguration is one way to achieve such a flexibility and then improving functionality and efficiency of existing business process. In [1], [2] and [3], the reconfiguration issue has been deeply investigated from both the theoretical and practical point of view. In this series of work modelling, analysis and implementation have been developed and discusses for a specific case study of workflow reconfiguration. In [4], instead, a synopsis of formalisms is presented and, according to specific criteria, the formalisms have been evaluated for their suitability to model dynamic reconfiguration.

As detailed in the papers mentioned above, once the reconfiguration of a business process has been performed, some (or all) of the system requirements might not hold anymore. That is why we will use a verification technique called model checking. Model checking is useful to verify correctness since it provides ways for simulating and testing a systems with data structures which can assume many different values. Furthermore, if the requirements are not met in the reconfiguration, model checking will produce a counterexample which can be used to pinpoint the source of the error.

The goal of our research is to develop a formal method for verifying the correctness of a reconfigurable business model. This purpose is achieved by developing a procedure that consists of several steps. We will now explain our approach and the procedure we developed. For further details on what a formal method is, please consult [5] and [6].

## II. Our Approach

Some verification techniques for business process models have been already presented in the past. For example, in [7] the behavioral specification of an application problem has been analyzed by using model checking. In [8], the same technique hs been used to detect potential information leakage in a business model. The differences between these works and ours are as follows. First, in this study we investigate the reconfiguration issue in BPMN model. Second, we intend to develop an algorithm that will translate the reconfigured BPMN model into a Promela model and, by using a model checking tool, verify if the properties are valid in this reconfigured model. For this purpose we introduce a formal specification of the BPMN model as an intermediate language. Fig.1 summarizes our verification procedure.

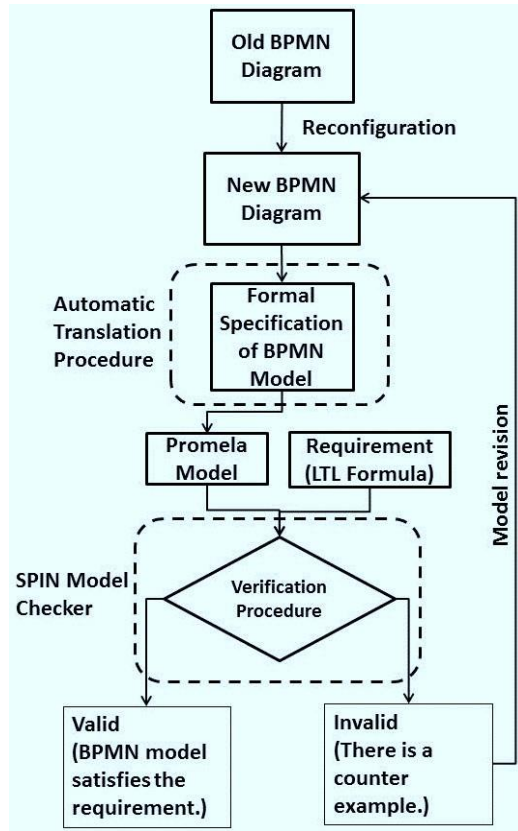

Figure 1. Procedure of the Verification

As for the first part of the procedure, we have the original (old) configuration of our business process defined as a BPMN diagram. After we have introduced the concept of reconfiguration, we achieve our reconfigured (new) BPMN diagram. This new model then needs to be explained by a formal specification. This formal specification will show the elements necessary to apply the automatic translation procedure which will generate a Promela model. This Promela model is necessary for the verification procedure together with the system requirements expressed in a LTL formulas. We will make use of the SPIN model checker as a tool for verifying the model, where the model specified in Promela is the one input and one of the system requirements, such as deadlock freeness, security, and reachability, is the other input. The output of SPIN model checker after the verification will show us the validation of the requirement property for the model. If the property is valid, it means that the reconfigured BPMN model satisfies the system requirement. If the property is invalid, the SPIN model checker will show us a counterexample that points out where did the requirement failed and finally we make a revision so that the model can satisfy the requirement.

Regarding the translation procedure, we have found that this idea has been investigated by [9] as a translation method to generate Promela models from UML design models. Also, we take into account the previous research in where there is a translation mapping procedure from BPMN into BPEL [10] and then taking the work done in [7] where shows an informal procedure of translating BPEL into Promela. However, these previous works do not provide a formal specification of the model as well as an algorithm for the translation. As for the system requirements, we can find many issues that will affect the correctness of our reconfigured model. Some of those issues are, for example, deadlock freedom, security, completeness (i,e,. the flow that starts in a state, must finish in another state) and any other specific property that the reconfiguration might have affected.

III. CONCLUSIONS AND FUTURE WORK

We presented a verification procedure for the reconfiguration of a BPMN model in a workflow environment. This procedure verifies if the requirements of a model, after introducing the reconfiguration, are valid. The verification procedure provides an automatic translation algorithm that is capable of generating readable Promela code by identifying the BPMN specification elements in a model.

As future work, we are interested in developing a fully automated verification procedure that will update the work done in the first stages of the procedure (reconfiguring the BPMN model) and will include GUI for the user.